\documentstyle[12pt]{article}
\begin{document}
\begin{titlepage}
\begin{flushright}
Z\"urich University ZU-TH 18/94\\
Pavia University FNT-T 94/21\\
\end{flushright}
\vfill
\begin{center}
{\large\bf A CASE FOR A BARYONIC DARK HALO}\\
\vfill
{\bf F. De Paolis$^{1,2,\#}$, G. Ingrosso$^{1,2,\#}$,
Ph.~Jetzer$^{3,}$* and M. Roncadelli$^4$}\\
\vskip 1.0cm
$^1$Dipartimento di Fisica, Universit\`a degli Studi di Lecce,
Via Arnesano CP 193,\\
 I-73100 Lecce, Italy,\\
$^2$INFN, Sezione di Lecce, Via Arnesano CP 193, I-73100 Lecce,
Italy,\\
$^3$Institute of Theoretical Physics, University of Z\"urich,
Winterthurerstrasse 190,\\
CH-8057 Z\"urich, Switzerland,\\
$^4$INFN, Sezione di Pavia, Via Bassi 6, I-27100 Pavia, Italy.
\end{center}
\vfill
\begin{center}
Abstract
\end{center}
\begin{quote}
Recent observations of microlensing events in the
Large Magellanic Cloud by the MACHO and EROS collaborations suggest
that an
important fraction of the galactic halo is in the form of Massive Halo
Objects (MHO) with mass $\sim 0.1 M_{\odot}$.
We outline a scenario in which dark clusters of MHO
with mass $\sim 0.1 M_{\odot}$ and H$_2$ molecular clouds
form in the halo at galactocentric distances larger than $\sim 10-20$
kpc,
provided baryons are a major constituent of the halo.
Possible signatures of our picture are discussed.

\end{quote}
\vfill
\begin{flushleft}
\vfill
\begin{center}
September 1994
\end{center}
\vskip 0.5cm
$^*$ Supported by the Swiss National Science Foundation.\\
$^{\#}$ Partially supported by Agenzia Spaziale Italiana.
\end{flushleft}
\end{titlepage}
\newpage

\baselineskip=21pt
One of the most important problems in astrophysics
concerns the nature of the dark
matter in galactic halos, as suggested by the observed
flat rotation curves.
Although various exotic dark matter candidates have
been proposed, present limits coming from primordial
nucleosynthesis allow a halo made of ordinary baryonic matter
that should be in the form of Massive Halo Objects (MHO) with masses in
the
range $10^{-7} < M/M_{\odot} < 10^{-1}$ \cite{kn:Derujula}.
Paczy\'nski suggested to detect MHO
using the gravitational lens effect \cite{kn:Paczynski}.

Recently, the French collaboration EROS \cite{kn:Aubourg} and the
American-Australian collaboration MACHO \cite{kn:Alcock} reported
the possible detection of altogether six microlensing events,
discovered by
monitoring over several years millions of stars in the Large Magellanic
Cloud (LMC), whereas the Polish-American collaboration OGLE
\cite{kn:Udalski}
and also the MACHO collaboration have found altogether at least forty
microlensing events by monitoring stars located in the galactic bulge.
Taking these results at face value an average mass
$\sim 0.1 M_{\odot}$ \cite{kn:Jetzer} for the MHO in the halo has been
derived.

At present the few events which have been found so far by monitoring
the
LMC do not allow to make a precise estimate of the fraction of dark
halo
matter in form of MHO, nor to infer whether the MHOs are located in the
halo or rather in the LMC itself \cite{kn:Gould1} or in a thick disk in
our
galaxy \cite{kn:Gould2}. It has also been proposed \cite{kn:Sahu} that
microlensing by stars within the LMC itself can account for the
observed
events.

Our aim is to present a scenario in which the halo of the Galaxy
substantially
consists of dark clusters of MHO and/or $H_2$ molecular clouds,
provided
baryons are a major constituent of the halo. We suggest that the dark
clusters
form at galactocentric distances larger than $\sim$ 10 - 20 kpc.

The proposed picture relies on the theory for the origin of stellar
globular clusters advocated by Fall and Rees \cite{kn:Fall}
and on the suggestion of Palla, Salpeter \&
Stahler \cite{kn:Palla} that the lower bound on the fragment masses in
a
collapsing, metal poor cloud can be as low as $10^{-2} M_{\odot}$.
Here we outline the main facts and for
more details see ref. \cite{kn:Depaolis}.

As a preliminary step, let us briefly recall the key-points of the
present
wisdom about the formation of stellar globular clusters. After its
initial
collapse, the proto Galaxy (PG) is expected to be shock heated to its
virial temperature $T_e \sim 10^6~K$. This temperature lies near a very
unstable region of the cooling curve \cite{kn:Dalgarno}, so that
density
enhancements should rapidly grow as the gas cools to lower
temperatures.
Fall and Rees \cite{kn:Fall} argued that overdense regions in the PG
cool more
rapidly than average (because the cooling rate by hydrogen
recombination is
proportional to $n^2_{H}$, where $n_{H}$ is the number density of
hydrogen)
and then a two phase medium forms with cool proto globular cluster
(PGC) clouds in pressure equilibrium with the external hot diffuse gas.
The PGC
clouds, which have temperature $T_c \sim 10^4~K$, are gravitationally
unstable
when their mass exceeds the Jeans mass
$M_J= 1.18~ (k_B T_c/\mu_c)^{2}~G^{-3/2}P_e^{-1/2}$,
where $\mu_c\sim 1.22~m_p$ is the mean molecular weight of the
primordial
neutral gas and $P_e$ is the pressure of the fully
ionized plasma that fills the PG.
Typical density of the PG is
$\rho_e\sim 1.7 \times 10^{-24}~(R/kpc)^{-1}~g~cm^{-3}$,
so that the resulting PGC cloud mass
and radius are $M_c \sim 10^{6}~(R/kpc)^{1/2}~M_{\odot}$ and
$r_c \sim (19~pc)~ (R/kpc)^{1/2} \Delta^{-1}$, where R is the
galactocentric
distance and $\Delta \sim 10$ is a factor which takes into account the
shrinking of the PGC cloud due to dissipation and collapse during its
subsequent evolution (for more details see \cite{kn:Carr}).
The main coolants below $10^4~K$ are molecular hydrogen and any heavy
element
produced by a first generation of stars. Molecular hydrogen, however,
would be
dissociated by various sources of UV radiation such as an active
galactic
nucleus (AGN) and/or a population of massive young stars in the
center of the PG. In the early more chaotic phase of the evolution
of the PG, in fact, an AGN is possibly formed at its center and/or a
first generation of massive stars due to the disruption of central PGC
clouds.
The last circumstance is realized
because, in the center of the PG, the
cloud collision time is shorter than the corresponding cooling time.
In a metal poor, dust free protogalactic gas
the molecular hydrogen abundance $f_{H_{2}}=n_{H_2}/n_H$ is determined
by
considering a set of atomic and molecular processes
which involve the creation of intermediate $H^-$ and $H_2^+$
\cite{kn:Shapiro}.
With the knowledge of $f_{H_2}$ one can compute the cooling rate
$\Lambda_c$
and the cooling time
\begin{equation}
\tau_{cool}=\frac{3\rho_c k_B T_c}{2\mu_c (\Lambda_c -\Gamma )}~,
\end{equation}
where the heating rate $\Gamma$ due to external radiation sources
has also been taken into account.
The subsequent evolution of the PGC cloud depends on the ratio between
$\tau_{cool}$ and the gravitational infall time
$\tau_{ff}=(3\pi/32G\rho_c)^{1/2}$ that, since $\rho_c \sim 400 \rho_e$
\cite{kn:Fall},
results to be $\tau_{ff} \sim 1.7 \times 10^6~(R/kpc)^{1/2}$ years.
If $\tau_{cool} \ll\tau_{ff}$ the PGC clouds rapidly cool to a
temperature
$T_c\sim 100~K$ before the gravitational instability sets in, while for
$\tau_{cool} \leq \tau_{ff}$
the PGC clouds cool and contract at the same time.
On the other hand,
when $\tau_{cool}\geq\tau_{ff}$ the PGC clouds do not cool,
remaining at a fixed temperature. The parameters that mainly
discriminate
between the different situations are $\Gamma$ and $\Lambda_c$.

In the inner halo, because of the presence of
an AGN and/or a first population of massive stars
at the center of the PG,
$H_2$ formation and cooling are heavily suppressed or delayed for a
wide
range of both external UV fluxes and/or PGC cloud densities
\cite{kn:Kang}.
For this case, in which $\tau_{cool}\geq\tau_{ff}$, the PGC clouds
remain
at $T_c\sim 10^4~K$ for a long time. This results
in an {\em imprinting} of a characteristic mass of
$M_c \sim 10^6~M_{\odot}$.
Moreover, during the permanence of the PGC clouds for a long
time in quasi hydrostatic equilibrium, propagation of sound waves
erases all large scale perturbations leaving only those
on small scale.
After enough $H_2$ has formed ($f_{H_{2}}\sim 10^{-3}$),
the temperature suddenly drops well below $10^4~K$
because now $\tau_{cool}\ll \tau_{ff}$.
The subsequent evolution of the PGC clouds goes on with a rapid growth
of the
small scale
perturbations that leads directly (in one step)
to the formation of stars within a narrow mass range \cite{kn:Murray}.
Thus, this scenario would explain
the formation of stellar globular clusters which are
observed today especially in the inner part of the galactic halo.
The formation of the PGC clouds could have been
delayed until after the Galaxy was
enriched by metals due to a first generation of stars, in this way
explaining
the observed absence of globular clusters with only primordial metal
abundances and the radial gradient of metallicity in the galactic
halo.

In the outermost part of the halo, where the incoming UV radiation flux
is suppressed due to the distance (so that $\tau_{cool}\leq\tau_{ff}$),
the
PGC clouds cool more gradually below $10^4~K$. Then, cooling and
collapse
occur simultaneously and their evolution proceeds according
to the scenario proposed by Palla et al. \cite{kn:Palla},
leading to a subsequent fragmentation into smaller clouds that remain
optically thin until the minimum value of the Jeans mass ($\leq 0.1
M_{\odot}$)
is attained.
In this case we expect that, because the PGC clouds do not
remain at a fixed temperature, there should be no imprinting of a
characteristic mass. In fact, when a PGC cloud is in a quiet ambient as
at the edge of the PG  the collapse
proceeds with a monotonic decrease of the Jeans mass
and a subsequent fragmentation
into clouds with lower and lower masses.
This process
stops when the fragments become optically thick to their own line
emission.
This happens because in a metal poor cloud virtually all the hydrogen
gas is
converted into molecular form due to three-body reactions
($H+H+H\rightarrow
H_2+H$ and $H+H+H_2\rightarrow 2H_{2}$). As a consequence of the
increased
cooling efficiency (due to the increase of $f_{H_{2}}$ up to 1) the
fragmentation process goes on until the Jeans mass reaches
$\sim 0.1~M_{\odot}$ or less, which is lower than the minimum mass for
nuclear
burning.

As a result of the above picture, dark clusters with MHO of mass
$\sim 0.1 M_{\odot}$ or less would form in the outer part of
the galactic halo. However, we don't expect the fragmentation process
to be
able to convert the whole gas mass contained in a PGC cloud into MHO
(see,
e.g., \cite{kn:Scalo}). Thus, we expect the remaining gas to form
self-gravitating $H_{2}$ molecular clouds \cite{kn:foot}, since in the
absence
of strong stellar winds (which in stellar globular clusters do eject
the gas)
the surviving gas remains gravitationally bound in the
dark cluster. The possibility that the gas is diffuse in the dark
cluster is
excluded due to its high virial temperature ($\sim 10^4 - 10^5$ K)
that would make the gas observable at 21 cm. In addition,
the gas cannot have diffused in the whole galactic halo because it
would have
been heated by the gravitational
field to a virial temperature $\sim 10^7~K$ and therefore would be
observable in the X-ray band (for which stringent upper limits are
available).
The further possibility that the gas entirely collapsed into the disk
is
also excluded because then its mass would be of the order of the
inferred dark halo mass.
Finally, we expect the ratio between the gas in the form of $H_2$
molecular
 clouds
and of MHO to be a function of the galactocentric distance which
determines the ambient conditions such as incoming UV fluxes
and collision rates among PGC clouds.

A few comments are in order. The UV flux incoming in PGC clouds is
crucial for
determining $f_{H_2}$ from which their subsequent evolution strongly
depends.
Since the formation of stellar globular clusters requires a
sufficiently high
UV flux, they can mainly form up to a certain galactocentric distance
$R_{crit}$ which we estimate
to be $\sim 10 - 20$  kpc \cite{kn:Depaolis}.
Beyond this distance the evolution of the PGC clouds gives rise to the
formation of dark clusters of MHO and/or $H_2$ molecular clouds.

A further question which naturally arises is whether dark clusters
are stable within the lifetime $t_g$ of the Galaxy. Standard
calculations
for the evaporation rate due to two-body encounters gives $t_{evap}
\sim 100~ t_{rel}$, where $t_{rel}$ is the usual two-body relaxation
time.
The condition $t_{evap} > t_g$ requires a constituent mass
$m \leq 10~ M_{\odot}$ \cite{kn:Carr}.
Another mechanism which could destroy dark clusters are collisions
among
themselves. We find that clusters are disrupted if they
are located within a certain galactocentric distance $R_{dis} \sim 10$
kpc
\cite{kn:Depaolis}.
{}From these considerations we conclude that dark clusters of MHO and/or
$H_2$
molecular clouds can still be present today at distances larger than
$R_{dis}$.

Let us briefly discuss the possible signatures of the above picture.
Besides
detection of microlensing events - which has in fact been
our main motivation - we should mention that
Maoz \cite{kn:Maoz} has recently considered the possibility to infer
from
microlensing observations whether MHO are clustered or not.
Already a relatively small number of events would
be sufficient to exclude this possibility, while to confirm it more
events are needed.

A further signature of the presence of $H_2$ molecular clouds in the
galactic
halo
should be a $\gamma$-ray flux produced through interaction with high
energy
cosmic ray protons. Cosmic rays scatter on protons in H$_2$ molecules
producing $\pi^0$'s which subsequently  decay into $\gamma$'s.

As a matter of fact, an
essential ingredient is the knowledge of the cosmic ray
flux in the halo. Unfortunately, this quantity is unknown
and the only information comes from theoretical estimates
\cite{kn:Breitschwerdt}. More precisely, from the mass-loss rate of a
typical galaxy, one can infer the total cosmic ray flux in the halo,
which turns out to be
$F\simeq 10^{41}~ A_{gal}^{-1} ~erg~cm^{-2}~ s^{-1}$, where $A_{gal}$
is the
surface of the galactic disk. Taking $R_{gal} \simeq 10$ kpc, we get
$F \simeq 1.1\times 10^{-4}~ erg~cm^{-2}~ s^{-1}$.
However, this  information
is not sufficient to carry out our calculations, since we need the
energy
distribution of the
cosmic rays.
We assume the same energy dependence
as measured on Earth and then scale the overall
density in such a way that the integrated energy flux agrees with the
above
value. Moreover, we assume that the cosmic ray density scales as
$1/R^2$ for large $R$ ($R$ is the galactocentric distance).
The measured primary cosmic ray flux on the Earth is
\begin{equation}
\Phi^{\oplus}_{CR}(E) \simeq \frac{2}{GeV}
\left(\frac{E}{GeV}\right)^{-2.7}
\frac{particles}{cm^2~ s~ sr}~, \label{eqno:42}
\end{equation}
and then the corresponding integrated energy flux is
$\simeq 5.7\times 10^{-2} ~erg~cm^{-2}~ s^{-1}$
(integration range:
$1~ GeV \leq E \leq 10^6~ GeV$).
Hence,
$\Phi_{CR}(E) \simeq 1.9\times 10^{-3}~\Phi_{CR}^{\oplus}$,
and then we get
\begin{equation}
\Phi_{CR}(E, R) \simeq \Phi_{CR}(E)~ \frac{a^2+R_{GC}^2}{a^2+R^2}~,
\label{eqno:45}
\end{equation}
where $a\sim 5$ kpc is the halo core radius and $R_{GC}\sim 8.5$ kpc is
our
distance from the galactic center. The source function
$q_{\gamma}(r)$ giving the photon number density at distance
$r$ from the Earth is
\begin{equation}
q_{\gamma}(r)= \sum_{n} \int dE_p~ dE_{\pi}~ \frac{4\pi}{c} \Phi_{CR}
(E_p,R(r))~ \frac{c\rho_{H_2}(R(r))}{m_p}~ \frac{d \sigma^n_{p
\rightarrow \pi}
(E_{\pi})}{dE_{\pi}}~ n_{\gamma}(E_p)~, \label{eqno:46}
\end{equation}
where $\sigma^n_{p \rightarrow \pi}(E_\pi)$ is the cross section for
the
reaction $pp \rightarrow n \pi^0$ ($n$ is the $\pi^0$ multiplicity),
$n_{\gamma}(E_p)$ is the photon multiplicity, $R(r)$ is the
galactocentric distance as a function of $r$,
while $\rho_{H_2}(R(r))$ is the fraction of dark
matter in form of H$_2$ molecular clouds (for which we assume the same
dependence on $R$ as in eq.(\ref{eqno:46})).
Actually, the cosmic ray protons in the halo
which originate from the galactic disk are mainly directed outwards.
This fact
implies that also the induced photons will leave the Galaxy.
However, the presence of magnetic fields in the halo might give rise
to a temporary confinement of the cosmic ray protons similarly to what
happens in the disk.
In addition, there could also be sources of
cosmic ray protons located in the halo itself, as for instance
isolated or binary pulsars in globular clusters.
Unfortunately, we are unable to give a quantitative estimate of the
above effects, so that we take them into account by introducing an
efficiency factor $\epsilon$, which could be rather small. In this way
the $\gamma$-ray photon flux reaching
the Earth is obtained by multiplying
$q_{\gamma}(r)$ by $\epsilon/4\pi r^2$ and integrating
the resulting quantity over the cloud volume along the line of sight.
The best chance, if any, to detect the $\gamma$-rays in question is
provided
by observations at high galactic latitude, and so we take
$\theta=90^0$.
Accordingly, we find for the $\gamma$-ray flux \cite{kn:Depaolis}
\begin{equation}
\Phi_{\gamma}(90^0) \simeq \epsilon~ 1.7 \times 10^{-6}~
\frac{photons}{cm^2~ s~sr}~. \label{eqno:53}
\end{equation}
The inferred upper bound for $\gamma$-rays in the 0.8 - 6 GeV
range for high galactic latitude is $3 \times 10^{-7}$ $photons~
cm^{-2}~
s^{-1}~ sr^{-1}$ \cite{kn:Bouquet}.
Hence, we see from eq. (\ref{eqno:53}) that the presence of
$H_2$ molecular clouds does not led at present to
any contradiction with the upper bound
provided $\epsilon < 10^{-1}$.

In conclusion, we have outlined a scenario for a baryonic dark halo
in which the formation of MHO and/or $H_2$ molecular clouds in the
outermost
part of the halo naturally arises, because in a quiet ambient the Jeans
mass
can reach values as low as $10^{-2}~ M_{\odot}$.
What is crucial
in discriminating the evolution of PGC clouds towards stellar globular
clusters or dark clusters are the decreasing collision rate and
UV flux with increasing galactocentric distances. The most promising
way
to detect dark clusters of MHO is through correlation effects in
microlensing observations, as we expect the dark clusters not to have
been disrupted up to now.
A much more difficult task is the detection of $H_2$ molecular clouds.
A possible signature of such clouds would be their $\gamma$-ray flux
induced
by halo cosmic ray protons.
Our calculation gives only an upper bound that is not in conflict with
present detection limits.
At any rate, the fact that our scenario naturally explains the
formation
of MHO at large galactocentric distances strongly supports - we believe
-
the idea that galactic dark matter is mostly baryonic, after all!

We would like to thank B. Bertotti, F. Palla and D. Pfenniger
for useful discussions.\\

\end{document}